\documentclass[prb,showpacs,nofootinbib,nofloats,endfloats,twocolumn]{revtex4}
\usepackage{graphicx}

\begin{document}


\title{Attractive Hubbard model with disorder and the generalized
Anderson theorem}

\author{E.Z. Kuchinskii$^1$, N.A. Kuleeva$^1$, M.V. Sadovskii$^{1,2}$}

\affiliation{$^1$Institute for Electrophysics, Russian Academy of Sciences,
Ural Branch, Ekaterinburg 620016, Russia\\
$^2$Institute for Metal Physics, Russian Academy of Sciences, 
Ural Branch, Ekaterinburg 620290, Russia}


\begin{abstract}

Using the generalized DMFT+$\Sigma$ approach we have studied disorder 
influence on single--particle properties of the normal phase and 
superconducting transition temperature in attractive Hubbard model.
The wide range of attractive potentials $U$ was studied --- from the
weak coupling region, where both the instability of the normal phase
and superconductivity are well described by BCS model, towards the
strong coupling region, where superconducting transition is due to
Bose -- Einstein condensation (BEC) of compact Cooper pairs, formed
at temperatures much higher than the temperature of superconducting
transition. We have studied two typical models of conduction band
with semi -- elliptic and flat densities of states, appropriate for 
three-dimensional and two-dimensional systems respectively.
For semi -- elliptic density of states disorder influence on all 
single-particle properties (e.g. density of states) is universal 
for arbitrary strength of electronic correlations and disorder and is 
due only to the general disorder widening of conduction band. 
In the case of flat density of states universality is absent in 
general case, but still the disorder influence is due mainly to 
band widening and universal behavior is restored for large enough disorder.
Using the combination of DMFT+$\Sigma$ and Nozieres -- Schmitt-Rink
approximations we have studied disorder influence upon superconducting 
transition temperature $T_c$ for the range of  characteristic 
values of $U$ and disorder, including the BCS-BEC crossover region and 
the limit of strong coupling. Disorder can either suppress $T_c$ 
(in the weak coupling region) or significantly increase $T_c$ 
(in strong coupling region). However in all cases the generalized 
Anderson theorem is valid and all changes of superconducting 
critical temperature are essentially due only to the general 
disorder widening of the conduction band.

\end{abstract}


\maketitle


\section{Introduction}

The problem of strong coupling superconductivity was studied for a long time, starting with
pioneering papers by Eagles and Leggett [\onlinecite{Eagles,Leggett}]. 
Significant progress here was achieved by Nozieres and Schmitt-Rink [\onlinecite{NS}], who suggested
an effective method to study the transition temperature crossover from weak coupling BCS-like
behavior towards Bose -- Einstein condensation (BEC) scenario in the strong coupling region.
Recent progress in experimental studies of quantum gases in magnetic and optical dipole traps,
as well as in optical lattices, with controllable parameters, such as density and interaction
strength (cf. reviews [\onlinecite{BEC1,BEC2}]), has increased the interest to superconductivity
(superfluidity of Fermions) with strong pairing interaction, including the region of BCS -- BEC
crossover. One of the simplest models allowing the study of BCS -- BEC crossover is the
Hubbard model with attractive on site interaction. The most successive approach to the solution
of Hubbard model, both in the case of repulsive interaction and for the studies of BCS -- BEC
crossover in case of attraction, is the dynamical mean field theory (DMFT)
[\onlinecite{pruschke,georges96,Vollh10}]. Attractive Hubbard model was studied within DMFT in a number of
recent papers [\onlinecite{Keller01,Toschi04,Bauer09,Koga11,JETP14}]. However, up to now there are only few
studies of disorder influence on the properties of normal and superconducting phases in this model,
especially in the region of BCS -- BEC crossover. Qualitatively disorder effects in this region
were analyzed in Ref. [\onlinecite{PosSad}], where it was argued, that Anderson theorem remains valid in
BCS -- BEC crossover region in the case of $s$-wave pairing. Diagrammatic approach to (weak) disorder
effects upon superconducting transition temperature and properties of the normal phase in crossover
region was developed recently in Ref. [\onlinecite{PalStr}]. 

In recent years we have developed the generalized DMFT+$\Sigma$ approach to Hubbard model
[\onlinecite{JTL05,PRB05,FNT06,UFN12}], which is very convenient for the studies of different external
(with respect to those taken into account in DMFT) interactions (such as pseudogap fluctuations
[\onlinecite{JTL05,PRB05,FNT06,UFN12}], disorder [\onlinecite{HubDis,HubDis2}], electron -- phonon interaction
[\onlinecite{e_ph_DMFT}]) etc. This approach is also well suited to analyze two--particle properties,
such as optical (dynamic) conductivity [\onlinecite{HubDis,PRB07}]. In Ref. [\onlinecite{JETP14}] 
we have used this approximation to calculate single -- particle properties of the normal phase
and optical conductivity in attractive Hubbard model. In a recent paper [\onlinecite{JTL14}] DMFT+$\Sigma$ 
approach was used by us to study disorder influence upon superconducting transition temperature, 
which was calculated in Nozieres -- Schmitt-Rink approximation. In this paper for the case of 
semi -- elliptic density of states of the ``bare'' conduction band, which is adequate for 
three -- dimensional systems,  we have numerically demonstrated the validity of the generalized
Anderson theorem, so that all the changes of critical temperature are controlled only by general
widening of the conduction band by disorder.

In this paper we present the analytic proof of such universal influence of disorder (in DMFT+$\Sigma$ 
approximation) upon single -- particle characteristics and temperature of superconducting transition
for the case of semi -- elliptic density of states, and also investigate disorder effects in the case
of the ``bare'' band with flat density of states, qualitatively appropriate for two -- dimensional systems.
It will be shown, that for the flat band model the universal dependence of single -- particle properties 
and superconducting transition temperature is also realized for the case of strong enough disorder.

\section{Disordered Hubbard model within DMFT+$\Sigma$ approach}

We consider the disordered nonmagnetic Hubbard model with attractive
interaction with Hamiltonian:
\begin{equation}
H=-t\sum_{\langle ij\rangle \sigma }a_{i\sigma }^{\dagger }a_{j\sigma
}+\sum_{i\sigma }\epsilon _{i}n_{i\sigma }-U\sum_{i}n_{i\uparrow
}n_{i\downarrow },  
\label{And_Hubb}
\end{equation}
where $t>0$ is transfer integral between nearest neighbors on the lattice, 
$U$ represents Hubbard -- like on site attraction, 
$a_{i\sigma }$($a_{i\sigma }^{\dagger}$) is annihilation (creation) operator 
of an electron with spin $\sigma$, $n_{i\sigma }=
a_{i\sigma }^{\dagger }a_{i\sigma }$ particle number operator on lattice site $i$, 
while local on site energies $\epsilon _{i}$ are assumed to be random variables 
(independent on different lattice sites). For the standard ``impurity'' diagram 
technique to be valid we take the Gaussian distribution of energy levels 
$\epsilon _{i}$:
\begin{equation}
\mathcal{P}(\epsilon _{i})=\frac{1}{\sqrt{2\pi}\Delta}\exp\left(
-\frac{\epsilon_{i}^2}{2\Delta^2}
\right)
\label{Gauss}
\end{equation}
Parameter $\Delta$ is the measure of disorder strength, while the Gaussian random field 
of random on site energy levels (independent on different sites -- ``white noise''
correlation) induces ``impurity'' scattering, which is analyzed using the standard
formalism of averaged Green's functions [\onlinecite{Diagr}].

The generalized DMFT+$\Sigma$ approach [\onlinecite{JTL05,PRB05,FNT06,UFN12}] extends the standard
dynamical mean field theory (DMFT) [\onlinecite{pruschke,georges96,Vollh10}]  taking into account
an additional ``external'' self-energy part  $\Sigma_{\bf p}(\varepsilon)$ 
(in general case momentum dependent), which is due to some additional interaction outside DMFT,
and gives an effective method to calculate both single -- particle and two -- particle
properties [\onlinecite{HubDis,PRB07}]. The success of this generalized approach is based upon the
choice of the single -- particle Green's function in the following form:
\begin{equation}
G(\varepsilon,{\bf p})=\frac{1}{\varepsilon+\mu-\varepsilon({\bf p})-\Sigma(\varepsilon)
-\Sigma_{\bf p}(\varepsilon)},
\label{Gk}
\end{equation}
where $\varepsilon({\bf p})$ is the  ``bare'' electronic dispersion, 
while the complete self -- energy is assumed to an additive sum of the local
self -- energy of DMFT and some ``external'' self -- energy $\Sigma_{\bf p}(\varepsilon)$,
due to neglect of the interference of Hubbard and ``external'' interactions.
This allows the conservation of the standard form of self -- consistent equations of the
standard DMFT [\onlinecite{pruschke,georges96,Vollh10}]. At the same time, at each step of DMFT iterations
we consistently recalculate an ``external'' self -- energy  $\Sigma_{\bf p}(\varepsilon)$ 
using the appropriate approximate scheme, corresponding to the form of an additional
interaction, while the local Green's function is also ``dressed'' by $\Sigma_{\bf p}(\varepsilon)$ 
at each step of the standard DMFT procedure. 

For ``external'' self -- energy entering  DMFT+$\Sigma$ cycle for the problem of random scattering
by disorder we use the simplest self -- consistent Born approximation, neglecting
diagrams with crossing ``impurity'' lines, which gives: 
\begin{equation}
\Sigma_{\bf p}(\varepsilon)\to\tilde\Sigma(\varepsilon)=\Delta^2\sum_{\bf p}G(\varepsilon,{\bf p}),
\label{BornSigma}
\end{equation}
where $G(\varepsilon,{\bf p})$ is the single -- electron Green's function (\ref{Gk}) and $\Delta$ is
the amplitude of site disorder.

To solve the effective single Anderson impurity problem of DMFT we used the numerical renormalization
group approach (NRG) [\onlinecite{NRGrev}].

In the following we consider two models of ``bare'' conduction band. The first one is the band with
semi -- elliptic density of states (per unit cell and single spin projection):
\begin{equation}
N_0(\varepsilon)=\frac{2}{\pi D^2}\sqrt{D^2-\varepsilon^2}
\label{DOSd3}
\end{equation}
where $D$ defined the band half-width. This model is appropriate for three -- dimensional system.
The second one is the model with the flat density of states, appropriate for two -- dimensional
case:
\begin{equation}
N_0(\varepsilon)=
\left\{
\begin{array}{ll}
\frac{1}{2D} & \quad |\varepsilon|\leq D\\ 
0 & \quad |\varepsilon| > D
\end{array}.
\right.
\label{DOSd2}
\end{equation}
In principle, for two -- dimensional systems we should take into account the presence of the
weak (logarithmic) Van Hove singularity in the density of states. However, this singularity
is effectively suppressed even by rather small disorder, so that the simple model of
Eq. \ref{DOSd2} is quite sufficient for our aims.

All calculations in this work has been done for the case of quarter -- filled band (the number of
electrons per lattice site n=0.5).

The temperature of superconducting transition in attractive model was analyzed in a number of papers
[\onlinecite{Keller01,Toschi04,Koga11}], both from the condition of instability of the normal phase  [\onlinecite{Keller01}] 
(divergence of Cooper susceptibility) and from the condition of superconducting order parameter going
to zero [\onlinecite{Toschi04,Koga11}]. In a recent paper [\onlinecite{JETP14}] we have determined the critical temperature
from the condition of instability of the normal phase, reflected in the instability of DMFT iteration
procedure. The results obtained in this way in fact coincided with those of Refs. [\onlinecite{Keller01,Toschi04,Koga11}]. 
Also in Ref. [\onlinecite{JETP14}] to calculate $T_c$ we have used the approach due to Nozieres and Schmitt-Rink
[\onlinecite{NS}], which allows the correct (though approximate) description of $T_c$ in BCS -- BEC crossover region.
In a later work [\onlinecite{JTL14}] we have used the combination of Nozieres and Schmitt-Rink and DMFT+$\Sigma$  
approximations for the detailed numerical studies of disorder dependence of $T_c$ and the number of local
pairs in the model with semi -- elliptic density of states.

\section{Disorder influence on single -- particle properties for the case of
semi--elliptic density of states}

In this section we shall analytically demonstrate, that in DMFT+$\Sigma$ approximation
disorder influence upon single -- particle properties of disordered Hubbard model (both attractive or
repulsive) with semi -- elliptic ``bare'' conduction band is completely described by effects of
general band widening by disorder scattering.

In the system of self -- consistent equations DMFT+$\Sigma$ equations [\onlinecite{PRB05,UFN12,HubDis}] both
information on the ``bare'' band and disorder scattering enter only on the stage of calculations of
the local Green's function:
\begin{equation}
G_{ii}=\sum_{\bf p}G(\varepsilon,{\bf p}),
\label{Gii_det}
\end{equation}
where the full Green's function $G(\varepsilon,{\bf p})$ is determined by Eq. (\ref{Gk}), while
the self -- energy due to disorder, in self -- consistent Born approximation, is defined by
Eq. (\ref{BornSigma}). Thus, the local Green's function takes the form:
\begin{eqnarray}
G_{ii}=\int_{-D}^{D}d\varepsilon' \frac{N_{0}(\varepsilon')}{\varepsilon+\mu-
\varepsilon'-\Sigma(\varepsilon)
-\Delta^2G_{ii}}=\nonumber\\
=\int_{-D}^{D}d\varepsilon' \frac{N_{0}(\varepsilon')}
{E_t-\varepsilon'},
\label{Gii_full}
\end{eqnarray}
where we have introduced the notation $E_t=\varepsilon+\mu-\Sigma(\varepsilon)-\Delta^2G_{ii}$.
In the case of semi -- elliptic density of states (\ref{DOSd3}) this integral is easily
calculated in analytic form, so that the local Green's function is written as: 
\begin{equation}
G_{ii}=2\frac{E_t-\sqrt{E_t^2-D^2}}{D^2}.
\label{Gii1}
\end{equation}
It is easily seen that Eq. (\ref{Gii1}) represents one of the roots of quadratic equation:
\begin{equation}
G_{ii}^{-1}=E_t-\frac{D^2}{4}G_{ii},
\label{Gii_eq}
\end{equation}
corresponding to the correct limit of $G_{ii}\to E_t^{-1}$ for infinitely narrow ($D\to 0$) band.
Then
\begin{eqnarray}
G_{ii}^{-1}=\varepsilon+\mu-\Sigma(\varepsilon)-\Delta^2G_{ii}-\frac{D^2}{4}G_{ii}=\nonumber\\
=\varepsilon+\mu-\Sigma(\varepsilon)-\frac{D_{eff}^2}{4}G_{ii},
\label{Gii_eq_eff}
\end{eqnarray}
where we have introduced $D_{eff}$ -- an effective half-width of the band (in the absence of
electronic correlations, i.e. for $U=0$) widened by disorder scattering:
\begin{equation}
D_{eff}=D\sqrt{1+4\frac{\Delta^2}{D^2}}.
\label{Deff}
\end{equation}
Eq. (\ref{Gii_eq}) was obtained from (\ref{Gii_full}), thus comparing (\ref{Gii_eq_eff}) and 
(\ref{Gii_eq}), we obtain: 
\begin{equation}
G_{ii}=\int_{-D_{eff}}^{D_{eff}}d\varepsilon' \frac{\tilde N_{0}(\varepsilon')}
{\varepsilon+\mu-\varepsilon'-\Sigma(\varepsilon)},
\label{Gii_full_eff}
\end{equation}
Here 
\begin{equation}
\tilde N_{0}(\varepsilon)=\frac{2}{\pi D_{eff}^2}\sqrt{D_{eff}^2-\varepsilon^2}
\label{}
\end{equation}
represents the density of states in the absence of interaction $U$
``dressed'' by disorder. This density of states remains semi -- elliptic in the 
presence of disorder, so that all effects of disorder scattering on single -- 
particle properties of disordered Hubbard model in DMFT+$\Sigma$ approximation 
are reduced only to disorder widening of conduction band, 
i.e. to the replacement $D\to D_{eff}$.

\section{Disorder influence on superconducting transition temperature}

Temperature of superconducting transition $T_c$ is not a single -- particle
characteristic of the system. Cooper instability, determining $T_c$ is 
related to divergence of two -- particle loop in Cooper channel. In the weak
coupling limit, when superconductivity is due to the appearance of Cooper pairs 
at $T_c$, disorder only slightly influences superconductivity with
$s$-wave pairing [\onlinecite{SCLoc,Genn}]. The so called Anderson theorem is valid and
changes of $T_c$ are connected only with the relatively small changes of the 
density of states by disorder. Th standard derivation of Anderson theorem
[\onlinecite{SCLoc,Genn}] uses the formalism of exact eigenstates of an electron in
the random field of impurities. Here we present another derivation of
Anderson theorem, using the exact Ward identity, which allows us to derive
the equation for $T_c$, which will be used to calculate $T_c$ in
Nozieres -- Schmitt-Rink approximation in disordered system.

In general, Nozieres -- Schmitt-Rink approach [\onlinecite{NS}] assumes, that corrections
due to strong pairing attraction significantly change the chemical potential of
the system, while possible correction due to this interaction to Cooper
instability condition can be neglected, so that we can always use here the weak
coupling (ladder) approximation. In such approximation the condition of Cooper
instability in disordered Hubbard model takes the form:
\begin{equation}
1=U\chi_0(q=0,\omega_m=0)
\label{Cupper}
\end{equation}
where
\begin{equation}
\chi_0(q=0,\omega_m=0)=T\sum_{n}\sum_{\bf pp'}\Phi_{\bf pp'}(\varepsilon_n)
\label{chi}
\end{equation}
represents the two -- particle loop (susceptibility) in Cooper channel ``dressed'' only by 
disorder scattering, and $\Phi_{\bf pp'}(\varepsilon_n)$ is the averaged two -- particle 
Green's function in Cooper channel  ($\omega_m=2\pi mT$ and $\varepsilon_n=\pi T(2n+1)$ 
are the usual Boson and Fermion Matsubara frequencies).

To obtain $\sum_{\bf pp'}\Phi_{\bf pp'}(\varepsilon_n)$ we use the exact Ward identity, 
derived by us in Ref. [\onlinecite{PRB07}]:
\begin{eqnarray}
G(\varepsilon_n,{\bf p})-G(-\varepsilon_n,-{\bf p})=\nonumber\\
=-\sum_{\bf p'}\Phi_{\bf pp'}(\varepsilon_n)(G_0^{-1}(\varepsilon_n,{\bf p'})-G_0^{-1}(-\varepsilon_n,-{\bf p'})),
\label{Word}
\end{eqnarray}
Here $G(\varepsilon_n,{\bf p})$ is the impurity averaged (but not containing Hubbard interaction corrections!) 
single -- particle Green's function. Using the obvious symmetry
$\varepsilon({\bf p})=\varepsilon(-{\bf p})$ and $G(\varepsilon_n,-{\bf p})=G(\varepsilon_n,{\bf p})$, 
we obtain from the Ward identity (\ref{Word}):
\begin{equation}
\sum_{\bf pp'}\Phi_{\bf pp'}(\varepsilon_n)=
-\frac{\sum_{\bf p}G(\varepsilon_n,{\bf p})-\sum_{\bf p}G(-\varepsilon_n,{\bf p})}{2i\varepsilon_n},
\label{Phi}
\end{equation}
so that for Cooper susceptibility (\ref{chi}) we have:
\begin{eqnarray}
\chi_0(q=0,\omega_m=0)=\nonumber\\
=-T\sum_{n}\frac{\sum_{\bf p}G(\varepsilon_n,{\bf p})-\sum_{\bf p}G(-\varepsilon_n,{\bf p})}{2i\varepsilon_n}=
\nonumber\\
=-T\sum_{n}\frac{\sum_{\bf p}G(\varepsilon_n,{\bf p})}{i\varepsilon_n}.
\label{chi1}
\end{eqnarray}
Performing now the standard summation over Matsubara frequencies [\onlinecite{Diagr}], we obtain:
\begin{eqnarray}
\chi_0(q=0,\omega_m=0)=\nonumber\\=-\frac{1}{4\pi i}\int_{-\infty}^{\infty}d\varepsilon
\frac{\sum_{\bf p}G^R(\varepsilon,{\bf p})-\sum_{\bf p}G^A(\varepsilon,{\bf p})}{\varepsilon}th\frac{\varepsilon}{2T}=\nonumber\\
=\int_{-\infty}^{\infty}d\varepsilon\frac{\tilde N(\varepsilon)}{2\varepsilon}th\frac{\varepsilon}{2T},
\nonumber\\
\label{chi2}
\end{eqnarray}
where $\tilde N(\varepsilon)$ is the density of states ($U=0$)  ``dressed'' by disorder scattering. 
In Eq. (\ref{chi2}) the energy $\varepsilon$ is reckoned from the chemical potential and if we reckon it from
the center of conduction band we have to replace $\varepsilon\to\varepsilon -\mu$, so that the condition of
Cooper instability (\ref{Cupper}) leads to the following equation for $T_c$:
\begin{equation}
1=\frac{U}{2}\int_{-\infty}^{\infty}d\varepsilon \tilde N_0(\varepsilon)\frac{th\frac{\varepsilon -\mu}{2T_c}}{\varepsilon -\mu} ,
\label{BCS}
\end{equation}
where $\tilde N_0(\varepsilon)$ is again the density of states (calculated for $U=0$) ``dressed'' by disorder scattering.
At the same time, the chemical potential of the system at different values of $U$ and $\Delta$ should be determined
from DMFT+$\Sigma$ calculations, i.e. from the standard equation for the number of electrons (band-filling),
determined by Green's function given by Eq. (\ref{Gk}), which allows us to find $T_c$ for the wide range of
model parameters, including the BCS-BEC crossover and strong coupling regions, as well as for different levels
of disorder. This reflects the physical meaning of Nozieres -- Schmitt-Rink approximation --- in the weak coupling
region transition temperature is controlled by the equation for Cooper instability (\ref{BCS}), while in the
limit of strong coupling it is determined as the the temperature of BEC, controlled by chemical potential.
Thus, the joint solution of Eq. (\ref{BCS}) and equation for the chemical potential guarantees the correct
interpolation for $T_c$ through the region of BCS-BEC crossover. This approach gives the results for the
critical temperature, which are quantitatively close to exact results, obtained by direct numerical DMFT 
calculations [\onlinecite{JETP14}], but demands much less numerical efforts.

It should be stressed, that we have used the exact Ward identity, which allows the use of Eq. (\ref{BCS}) 
also in the region of strong disorder, when the effects of Anderson localization may become relevant.
Eq. (\ref{BCS}) demonstrates, that the critical temperature depends on disorder only through the disorder
dependence of the density of states $\tilde N(\varepsilon)$, which is the main statement of Anderson theorem.
In the framework of Nozieres -- Schmitt-Rink approach Eq. (\ref{BCS}) is conserved also in the region of
strong coupling, when the critical temperature is determined by BEC condition for compact Cooper pairs. 
In this case the chemical potential $\mu$, entering Eq. (\ref{BCS}),  may significantly depend on disorder.
However, in DMFT+$\Sigma$ approximation this dependence of chemical potential (as well as any other single --
particle characteristic) in the model with semi -- elliptic density of states is only due to disorder
widening of conduction band. Thus, both in BCS -- BEC crossover and strong coupling regions the generalized
Anderson theorem actually remains valid. Correspondingly, in the model of semi -- elliptic band Eq. (\ref{BCS})
leads to universal dependence of $T_c$ on disorder, due to the change of $D\to D_{eff}$. Such universality
is fully confirmed by numerical calculations of $T_c$ in this model, performed in Ref. [\onlinecite{JTL14}] 
(cf. also the results presented below).

\section{Main results}

\begin{figure*}
\center{\includegraphics[clip=true,width=0.6\textwidth]{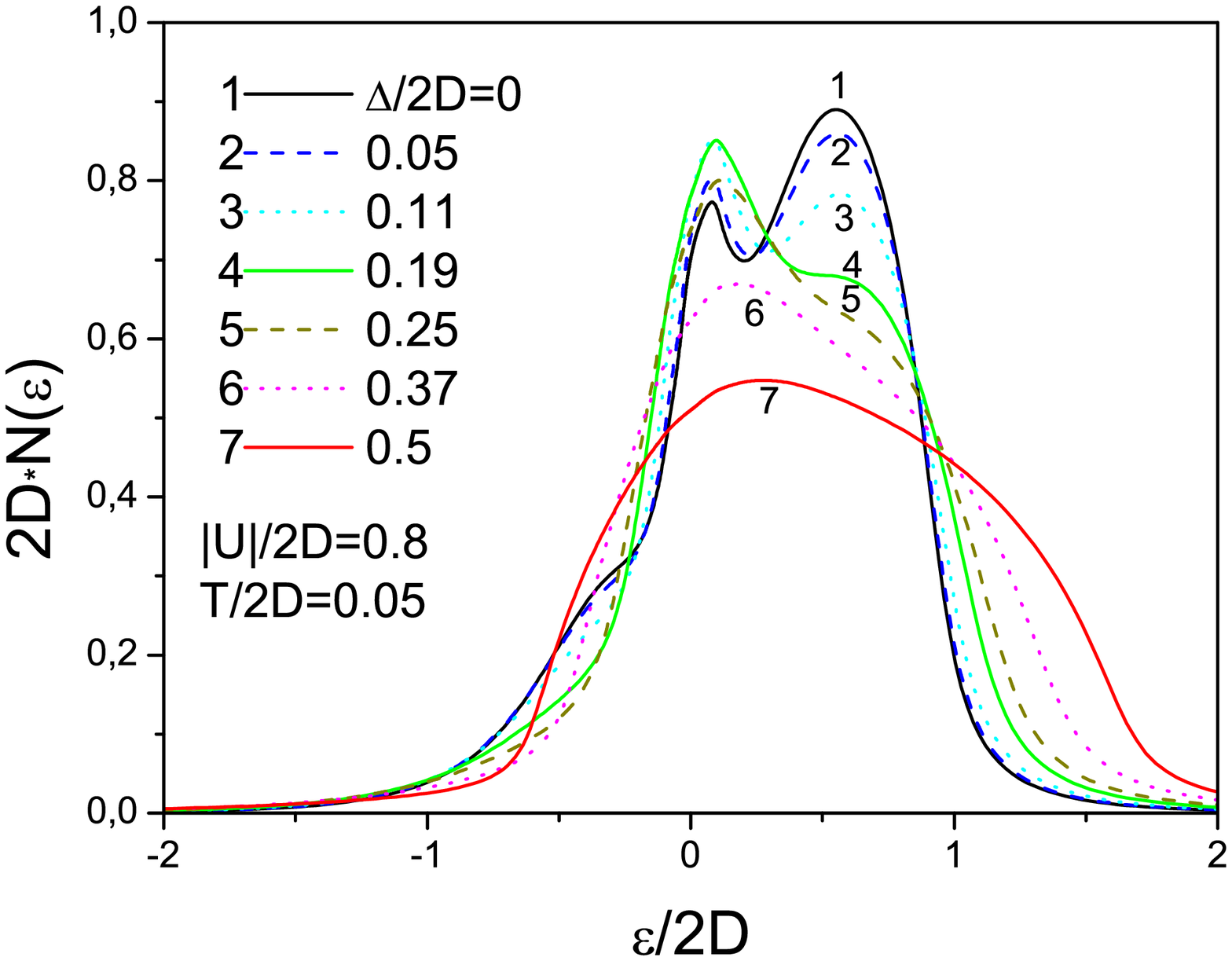}}
\caption{Dependence of the density of states on disorder in the model with
semi -- elliptic band.}
\label{fig1}
\end{figure*}

\begin{figure*}
\center{\includegraphics[clip=true,width=\textwidth]{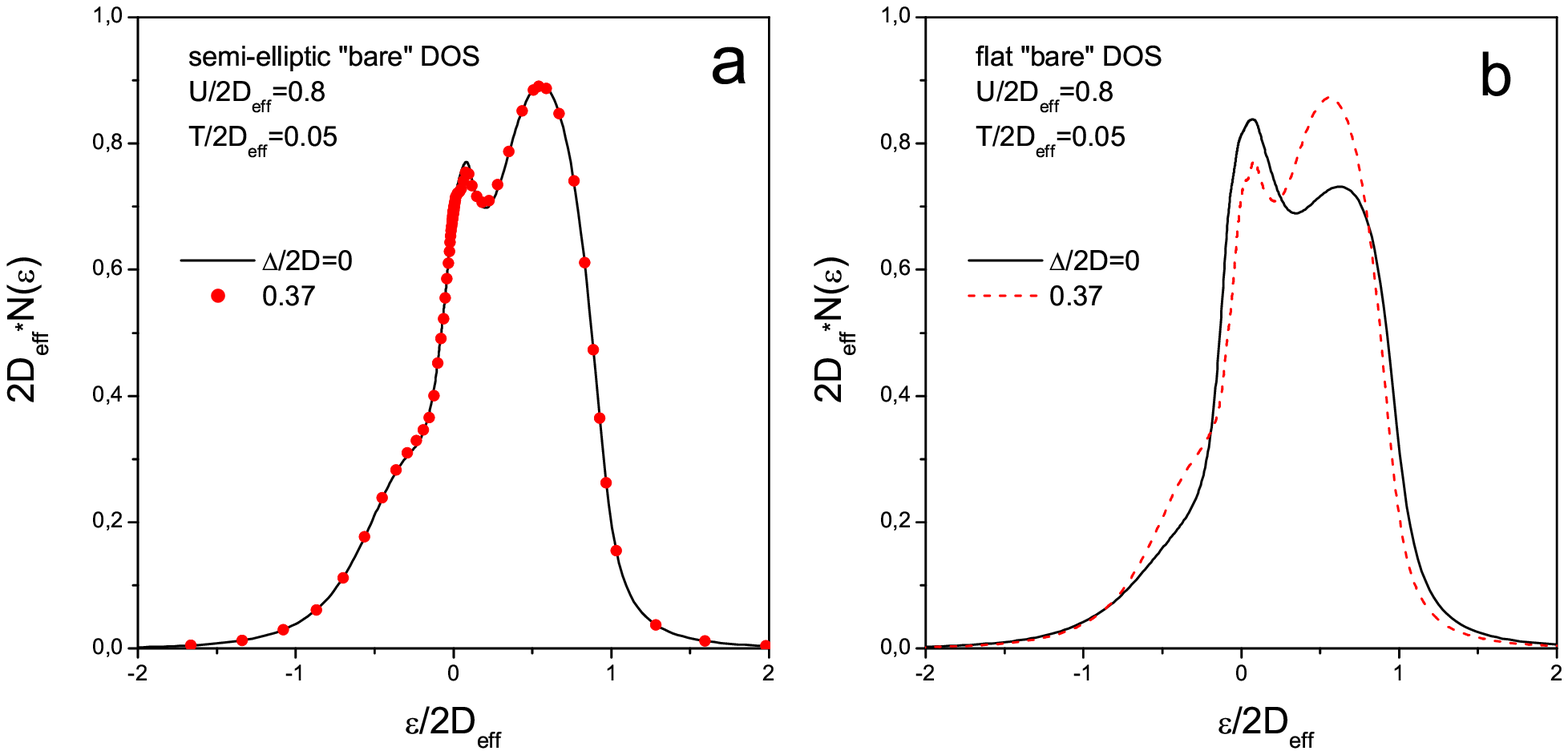}}
\caption{Universal dependence of the density of states on disorder: (a) ---  the model of semi -- elliptic ``bare'' density of states; 
(b) --- the model of flat ``bare'' density of states.}
\label{fig2}
\end{figure*}

Let us now discuss the main results of our numerical calculations, explicitly 
demonstrating the universal behavior of single -- particle properties and
superconducting transition temperature with disorder. We shall see, that all
disorder effects are effectively controlled, in fact, only by the growth of
half-width of conduction band, which for the case of semi -- elliptic density
of states are given by Eq. (\ref{Deff}). In case of the band with flat density
of states, the growth of disorder changes the shape of the density of states,
making it semi -- elliptic in the limit of strong enough disorder, while the
effective half-width of the band is given by (cf. Appendix A):
\begin{equation}
\frac{D_{eff}}{D}=\sqrt{1+\frac{\Delta^2}{D^2}}+\frac{1}{2}\frac{\Delta^2}{D^2}
\ln\left(\frac{\sqrt{1+\frac{\Delta^2}{D^2}}+1}{\sqrt{1+\frac{\Delta^2}{D^2}}-1}\right).
\label{Deffd2}
\end{equation}
As an example of the most important single -- particle property we take the
density of states. In Fig. \ref{fig1} we show the evolution of the density of
states with disorder in the model of semi -- elliptic band [\onlinecite{JETP14}].
We can see, that the growth of disorder smears the density of states and
widens the band. This smearing somehow masks the peculiarities of the density
of states due to correlation effects. In particular, both  quasiparticle peak and
lower and upper Hubbard bands, observed in Fig. \ref{fig1} in the absence of 
disorder are completely destroyed in the limit of strong enough disorder.
However, we can easily convince ourselves, that this evolution is only due to
the general widening of the band due to disorder (cf. (\ref{Deff}), (\ref{Deffd2})),
as all the data for the density of states belong to the same universal curve
replotted in appropriate new variables, with all energies (and temperature) normalized 
by the effective bandwidth by replacing $D\rightarrow D_{eff}$, as shown in 
Fig. \ref{fig2}(a), in complete accordance with the general results, obtained 
above. In the case of conduction band with flat density of states, there is no
complete universality, as is seen from Fig. \ref{fig2}(b) for low enough values of
disorder. However, for large enough disorders the dashed curve shown in 
Fig. \ref{fig2}(b) practically coincides with universal curve for the density of
states< shown in Fig. \ref{fig2}(a). This reflects the simple fact, that at large disorders
the flat density of states effectively transforms into semi -- elliptic (cf. Appendix A).

\begin{figure*}
\center{\includegraphics[clip=true,width=\textwidth]{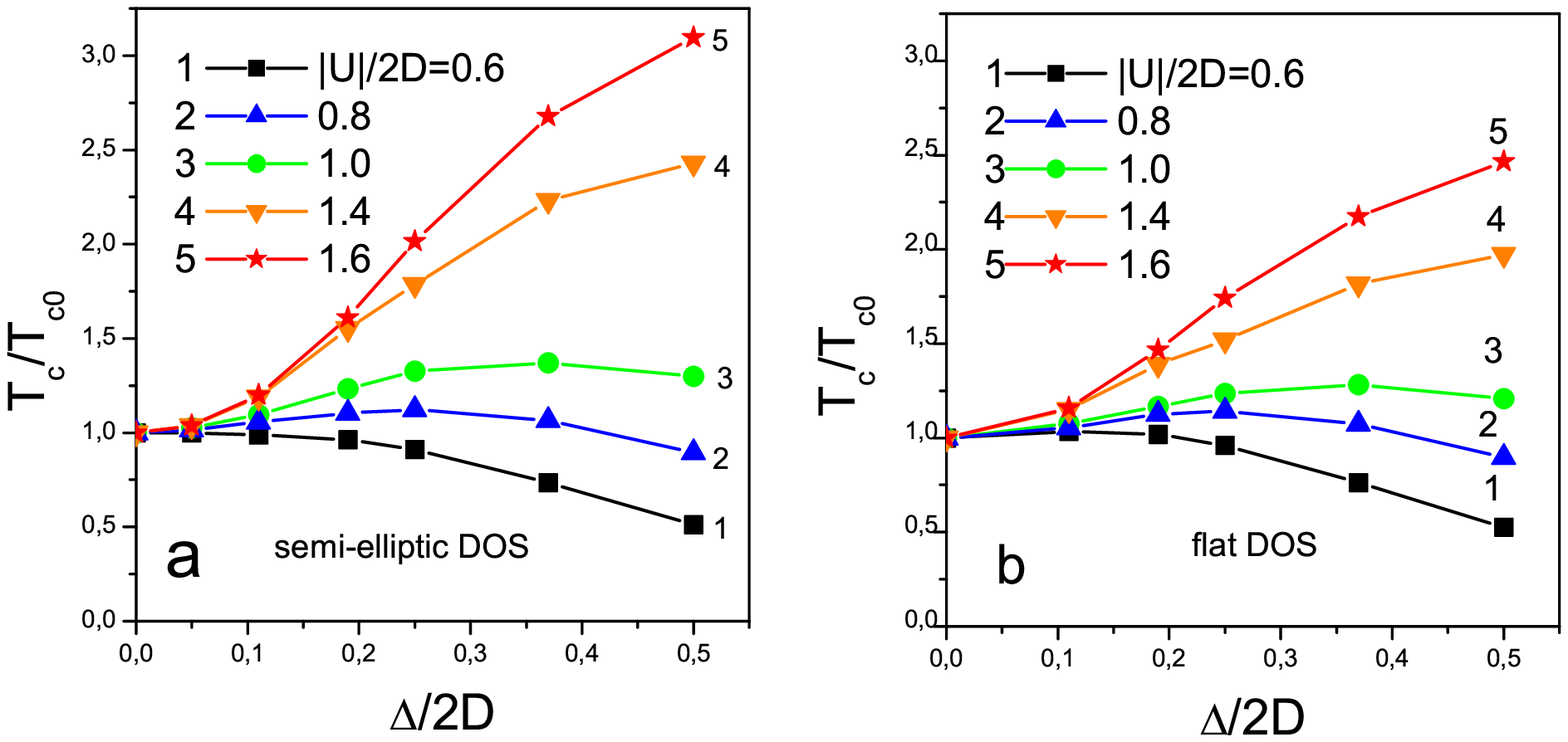}}
\caption{Dependence of superconducting transition temperature on disorder for different values of Hubbard
attraction $U$: (a) --- semi -- elliptic band; (b) --- flat band.}
\label{fig3}
\end{figure*}

\begin{figure*}
\center{\includegraphics[clip=true,width=\textwidth]{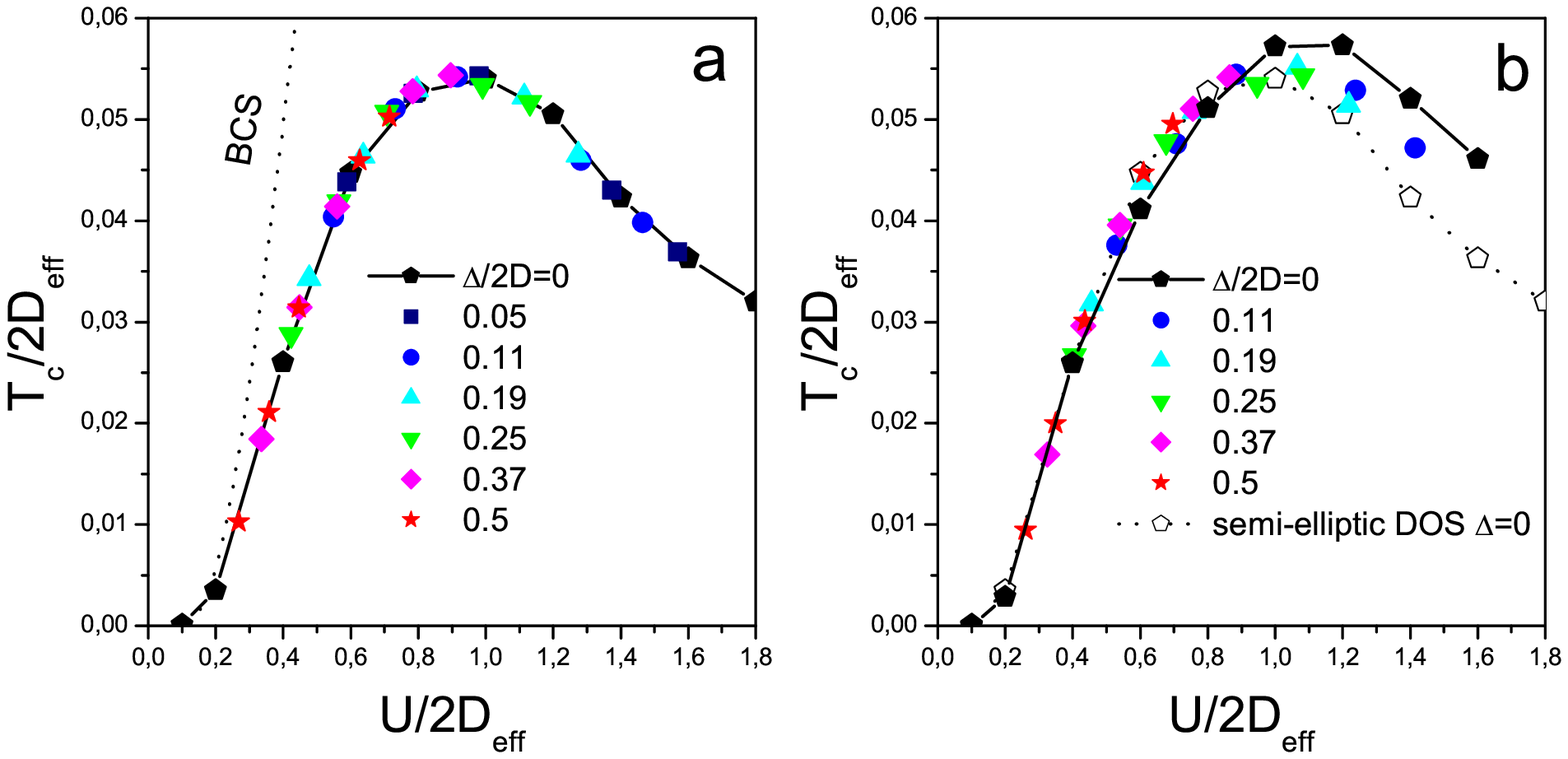}}
\caption{Universal dependence of superconducting critical temperature on Hubbard attraction $U$ for different
disorder levels: (a) --- semi -- elliptic band. Dashed curve represent BCS dependence in the absence of disorder.
(b) --- flat band. Dashed line represents similar dependence for semi -- elliptic band for $\Delta =0$.} 
\label{fig4}
\end{figure*}

Going now to the analysis of superconducting transition temperature, in Fig. \ref{fig3} we
present the dependence of $T_c$ (normalized by the critical temperature in the absence of
disorder $T_{c0}=T_c(\Delta=0$)) on disorder for different values of pairing interaction $U$
for both models of initial ``bare''  density of states (semi -- elliptic --- Fig.\ref{fig3}(a) 
and flat --- Fig. \ref{fig3}(b)). Qualitatively the evolution of $T_c$ with disorder is the same
for both models. We can see, that in the weak coupling limit ($U/2D\ll1$) disorder slightly
suppresses $T_c$ (curves 1). At intermediate couplings ($U/2D\sim 1$) weak disorder increases $T_c$, 
while the further growth of disorder suppresses the critical temperature (curves 3). 
In the strong coupling region ($U/2D\gg 1$) the growth of disorder leads to significant
increase of the critical temperature (curves 5). However, we can easily see, that such
complicated dependence of $T_c$ on disorder is completely determined by disorder
widening of the ``bare'' ($U=0$) conduction band, demonstrating the validity of the generalized
Anderson theorem for all values of $U$. In Fig. \ref{fig4} curve with octagons show the dependence
of the critical temperature $T_c/2D$ on coupling strength $U/2D$ in the absence of disorder
($\Delta =0$) for both model of ``bare'' conduction bands (semi -- elliptic --- Fig. \ref{fig4}(a) 
and flat --- Fig. \ref{fig4}(b)). We can see, that in both models in the weak coupling region
superconducting transition temperature is well described by BCS model (in Fig. \ref{fig4}(a) 
the dashed curve represents the result of BCS model, with $T_c$ defined by Eq. (\ref{BCS}), with
chemical potential independent of $U$ and determined by quarter -- filling of the ``bare'' band),
while in the strong coupling region the critical temperature is determined by BEC condition for
Cooper pairs and drops as $t^2/U$ with the growth of $U$ (inversely proportional to the effective mass of
the pair), passing through the maximum at $U/2D_{eff}\sim 1$. The other symbols in Fig. \ref{fig4}(a) 
show the results for $T_c$ obtained by combination of DMFT+$\Sigma$ and Nozieres -- Schmitt-Rink approximations
for the case of semi -- elliptic ``bare'' band. We can see, that all data (expressed in normalized
units of $U/2D_{eff}$ and $T_c/2D_{eff}$) ideally fit the universal curve, obtained in the absence
of disorder. For the case of flat ``bare'' band, results of our calculations are shown in
Fig. \ref{fig4}(b) and we do not observe the complete universality --- data points, corresponding to
different degrees of disorder somehow deviate from the curve, obtained in the absence of disorder. 
However, with the growth of disorder the form of the band becomes close to semi -- elliptic and our
data points move towards the universal curve, obtained for semi -- elliptic case and shown by the
dashed curve in Fig. \ref{fig4}(b), thus confirming the validity of the generalized Anderson theorem.

\section{Conclusion}

In this paper, in the framework of DMFT+$\Sigma$ generalization of dynamical mean field theory,
we have studied disorder influence on single -- particle properties (e.g. density of states)
and temperature of superconducting transition in attractive Hubbard model.
Calculations were made for a wide range of attractive interactions $U$, from the weak
coupling region of $U/2D_{eff}\ll 1$, where both instability of the normal phase and
superconductivity is well described by BCS model, up to the strong coupling limit of
$U/2D_{eff}\gg 1$, where superconducting transition is determined by Bose -- Einstein
condensation of compact Cooper pairs, forming at temperatures much higher than the
temperature of superconducting transition. We have shown analytically, that in the case
of conduction band with semi -- elliptic density of states, which is a good approximation
for three -- dimensional case, disorder influences all single -- particle properties in a
universal way --- all changes of these properties are due only to disorder widening of
the band. In the model of conduction band with flat density of states, which is appropriate
for two -- dimensional systems, there is no universality in the region of weak disorder.
However, the main effects are again due to general widening of the band and complete 
universality is restored for high enough disorders, when the density of states effectively
becomes semi -- elliptic.

To study the superconducting transition temperature we have used the combination of
DMFT+$\Sigma$ approach and Nozieres --- Schmitt-Rink approximation. For both models of
conduction band density of states disordering may either suppress the critical temperature
$T_c$ (in the region of weak coupling) or significantly increase it (in the strong coupling
region). However, in all these cases we have actually proven the validity of the generalized
Anderson theorem. so that all the changes of transition temperature are, in fact, controlled
only by the effects of general disorder widening of the conduction band.
In case of initial semi -- elliptic band disorder influence on $T_c$ is completely universal,
while in the case of initial flat band such universality is absent at weak disorder, but is
completely restored for high enough disorder levels.

Finally we should like to present some additional comments on the methods and
approximations used. Both DMFT+$\Sigma$ and Nozieres -- Schmitt-Rink
approaches represent cetrain approximate interpolation schemes, strictly
valid only in corresponding limiting cases (e.g. small disorder or small (large)
$U$). However, both schemes demonstrate their effectiveness also in the case of 
intermediate values of $U$ and intermediate (or even strong) disorder.
Actually, the effectiveness of Nozieres -- Schmitt-Rink (neglecting $U$ 
corrections in Cooper channel) approximation was verified by comparison with 
direct DMFT calculations [\onlinecite{JETP14}].
The use of DMFT+$\Sigma$ to analyze the disorder effects in repulsive Hubbard 
model was shown to produce reasonable results for the phase diagram, as
compared to exact numerical simulations of disorder in DMFT, including the
region of large disorder (Anderson localized phase) 
[\onlinecite{UFN12,HubDis,HubDis2}]. However, the role of approximations made 
in DMFT+$\Sigma$, such as the neglect of the intrference of disorder
scattering and correlation effects, deserves further studies.

This work is supported by RSF grant No. 14-12-00502.


\appendix

\section{}

For the band with flat density of states (at $U=0$ and $\Delta =0$) disorder leads both
to widening of the band and to the change of the form of the density of states. 
Taking the density of states in the form given by Eq. (\ref{DOSd2}) we calculate the local
Green's function as:
\begin{eqnarray}
G_{ii}=\frac{1}{2D}\int_{-D}^{D}d\varepsilon' \frac{1}
{\varepsilon-\varepsilon'-\Delta^2G_{ii}}=\nonumber\\
\frac{1}{2D}\ln\left(\frac{\varepsilon-\Delta^2G_{ii}+D}
{\varepsilon-\Delta^2G_{ii}-D}\right),
\label{Gii_flat}
\end{eqnarray}
where energy $\varepsilon$ is reckoned from the middle of the ``bare'' band. 
Let us introduce auxiliary notations, writing $G_{ii}=R-iI$. 
At the band edges $I\to 0$, so that expanding the r.h.s. of Eq. (\ref{Gii_flat}) 
up to linear terms in $I$, we get:
\begin{equation}
R-iI\approx\frac{1}{2D}\ln\left(\frac{\varepsilon-\Delta^2R+D}
{\varepsilon-\Delta^2R-D}\right)-
iI\frac{\Delta^2}{(\varepsilon-\Delta^2R)^2-D^2}
\label{Gii_flat_eq}
\end{equation}
Equating the real parts in (\ref{Gii_flat_eq}) we obtain
$R=\frac{1}{2D}\ln\left(\frac{\varepsilon-\Delta^2R+D}
{\varepsilon-\Delta^2R-D}\right)$. 
Similarly, equating the imaginary parts at the band edges we get
$\varepsilon-\Delta^2R=\pm\sqrt{D^2+\Delta^2}$, and substituting this 
expression into logarithm in the previous expression, we find $R$ 
and band edges positions at:
\begin{equation}
\varepsilon=\pm\left(\sqrt{D^2+\Delta^2}+\frac{\Delta^2}{2D}
\ln\left(\frac{\sqrt{D^2+\Delta^2}+D}{\sqrt{D^2+\Delta^2}-D}\right)
\right)
\label{Deff_flat}
\end{equation}
Thus, the half-width of the band $D_{eff}$ widened by disorder in this model is
determined by Eq. (\ref{Deffd2}) used above.

We should note, that the Born approximation for disorder scattering used by us,
though formally valid only for small disorder $\Delta\ll D$, the effects of Anderson
localization at large disorders $\Delta\sim D$ do not qualitatively change the
density of states [\onlinecite{SCLoc}], so that Born approximation gives
qualitatively correct results also in the region of large disorder. Actually,
this approximation neglects only the appearance  exponentially small ``tails'' in the
density of states, outside the ``mean field'' band edges [\onlinecite{SCLoc}] and
gives more or less correct results inside such a band.

At large enough disorders almost any  ``bare'' band width bandwidth $2D$ and
arbitrary density of states $N_{0}(\varepsilon)$ acquires semi -- elliptic density
of states. In the limit of very large disorder $\Delta\gg D$ almost in the whole
band, widened by disorder, we have $|\varepsilon-\Delta^2R|\gg D$ and in the
expression for the local Green's function we can neglect $\varepsilon'$-dependence
in the denominator of the integrand:  
\begin{equation}
R-iI=G_{ii}=\int_{-\infty}^{\infty}d\varepsilon' \frac{N_{0}(\varepsilon')}
{\varepsilon-\varepsilon'-\Delta^2G_{ii}}
\approx\frac{1}{\varepsilon-\Delta^2R+i\Delta^2I}
\label{Gii_lim}
\end{equation}
Then we immediately get:
\begin{equation}
\varepsilon-\Delta^2R=\frac{\varepsilon}{2};
\qquad I=\frac{1}{2\Delta^2}\sqrt{4\Delta^2-\varepsilon^2}
\label{RI_lim}
\end{equation}
so that the density of states ``dressed'' by disorder
\begin{equation}
N(\varepsilon)=-\frac{1}{\pi}ImG_{ii}=\frac{I}{\pi}=
\frac{2}{\pi(2\Delta)^2}\sqrt{(2\Delta)^2-\varepsilon^2}
\label{DOS_lim}
\end{equation}
becomes semi -- elliptic (\ref{DOSd3}) with half-width $D_{eff}=2\Delta$. 
Thus, at strong enough disorder any ``bare'' band becomes semi -- elliptic,
restoring universal dependence of single -- particle properties on
disorder discussed above. In this sense, the model of the ``bare'' band
with semi -- elliptic density of states is most appropriate for the studies
of the effects of strong disorder.


\newpage

\onecolumngrid

\newpage

\end{document}